\documentclass{article}%
\usepackage{amsmath}
\usepackage{amsfonts}
\usepackage{amssymb}
\usepackage{graphicx}%
\newcommand{\bpartial}{\mathop{\partial\kern -4pt\raisebox{.8pt}{$|$}}}
\newcommand{\bra}{\mathopen{[\kern-1.6pt[}}
\newcommand{\ket}{\mathclose{]\kern-1.5pt]}}
\newcommand{\bbra}{\mathopen{[\kern-2.2pt[\kern-2.3pt[}}
\newcommand{\bket}{\mathclose{]\kern-2.1pt]\kern-2.3pt]}}

\makeindex
\begin{document}
\title{\bf Perturbed Wess-Zumino-Witten models and N=(2,2) supersymmetric sigma models on Lie groups with complex structure}
\author { A. Rezaei-Aghdam \hspace{-3mm}{ \footnote{e-mail: rezaei-a@azaruniv.edu}\hspace{2mm} {\small and }
M. Sephid \hspace{-3mm}{ \footnote{ e-mail: s.sephid@azaruniv.edu}}\hspace{2mm}} \\
{\small{\em Department of Physics, Faculty of Sciences,  }}\\
{\small{\em Azarbaijan Shahid Madani University,}}\\
{\small{\em 53714-161, Tabriz, Iran  }}}
\maketitle
\begin{abstract}

We have perturbed Wess-Zumino-Witten (WZW)
models and also N=(2,2) supersymmetric sigma models on Lie groups by adding a term containing complex structure
to their actions. Then, using non-coordinate basis, we have shown
that for N=(2,2) supersymmetric sigma models on Lie groups the conditions (from the algebraic point of view) for the preservation of the N=(2,2) supersymmetry impose
that the complex structure must be invariant; so only the Abelian Lie algebras admit these deformations preserving the N=(2,2) supersymmetry. Also, we have shown that the perturbed WZW model with this term, using Hermitian (not necessarily invariant) condition, is an integrable model.
\end{abstract}
\newpage
\section{\bf Introduction}

Supersymmetric sigma models are of particular interest, for their
intimate connection to complex geometry of target manifold
\cite{Gat},\cite{sev} and for their role which is effective low-energy
actions for supergravity scalars. The N=(2,2) extended
supersymmetry in sigma model from the geometrical point of view is
equivalent to the existence of bi-Hermitian structure on the
target manifold where the complex structures are covariantly
constant with respect to torsionful affine connections \cite{Gat}
(see also \cite{Lya} and references therein). We know that the
algebraic structures related to bi-Hermitian relations of the
N=(2,2) supersymmetric WZW models are the Manin triples
\cite{Par}\cite{Getzler}\cite{Lin}. Furthermore, recently the algebraic
structure associated with the bi-Hermitian geometry of the N=(2,2)
supersymmetric sigma models on Lie groups has been found in \cite{RS}.
In \cite{ER}, we have studied the perturbed N=(2,2) supersymmetric
sigma models on Lie groups and have obtained conditions under
which the N=(2,2) supersymmetry is preserved. Here, in this
direction, we perturb N=(2,2) supersymmetric sigma models on Lie groups and also bosonic WZW
models by adding term containing complex structure
to their actions. We have shown that for N=(2,2) supersymmetric sigma models on Lie groups the preservation of the N=(2,2)
supersymmetry imposes that the invariant complex structure must be compatible with
ad-invariant metric, so from \cite{JO} (where it is proved that only the Abelian Lie algebras admit these structures); only the Abelian Lie algebras admit deformation which preserve the N=(2,2) supersymmetry. Also,
we have shown that the perturbed WZW model with complex structure is an
integrable model if and only if the complex structure be Hermitian. The paper is organized as follows.

In section 2, we first review the N=(2,2) supersymmetric sigma
models, then using the method mentioned in \cite{ER} we have shown
that the perturbed N=(2,2) supersymmetric sigma model on Lie group
has the N=(2,2) supersymmetry when the complex structure is an invariant, i.e., only the Abelian Lie algebras admit a deformation which preserve the N=(2,2) supersymmetry. In section 3, we
perturb WZW model with the bosonic version of the term which is
used in section 2. Then, we show that this perturbed model is an
integrable model when the tensor $J$ is a Hermitian complex structure (not necessarily invariant).
Finally, at the end of this section we present an example using the
Heisenberg Lie group. Some concluding remarks are addressed in
section 4.
\vspace{-.5cm}
\section{\bf \bf N=(2,2) supersymmetric sigma model on Lie groups perturbed with complex structure}

We know that the N=(2,2) supersymmetric sigma models \cite{Gat} on
the manifold M can be shown by the following N=1 supersymmetric sigma models
action
\begin{equation}\label{1}
S=\int d^{2}\sigma d^{2}\theta
D_{+}\Phi^{\mu}D_{-}\Phi^{\nu}(G_{\mu\nu}(\Phi)+B_{\mu\nu}(\Phi)),
\end{equation}
where $\Phi^{\mu}$ are N=1 superfields with bosonic parts as
coordinates of the manifold M such that the bosonic parts of
$G_{\mu\nu}$ and $B_{\mu\nu}$ are, respectively, the metric and the
antisymmetric tensors on M. This action is manifestly invariant
under supersymmetry transformations
\begin{equation}\label{2}
\delta^{1}(\epsilon)\Phi^{\mu}=i(\epsilon^{+}Q_{+}+\epsilon^{-}Q_{-})\Phi^{\mu},
\end{equation}
furthermore, it has additional non-manifest supersymmetry of the
form
\begin{equation}\label{3}
\delta^{2}(\epsilon)\Phi^{\mu}=\epsilon^{+}D_{+}\Phi^{\nu}J^{\mu}_{+\nu}(\Phi)+\epsilon^{-}D_{-}\Phi^{\nu}J^{\mu}_{-\nu}(\Phi),
\end{equation}
where in the above relations $Q_{\pm}$ and $D_{\pm}$ are
supersymmetry generators and superderivatives, respectively, and
$J^{\rho}_{\pm\sigma} \in TM \otimes T^{\ast}M$ are complex
structures. Invariance of the action (\ref{1}) under the
transformations (\ref{3}) imposes the fact that $J_\pm$ must be
bi-Hermitian complex structure such that their covariant
derivatives with respect to torsionful affine connections
($\Omega^{\pm\mu}_{\rho\nu}=\Gamma^{\mu}_{\rho\nu}\pm
G^{\mu\sigma}H_{\sigma\rho\nu})$\hspace{-1mm}{ \footnote{Note that
$H$ is the torsion three form
$H_{\mu\rho\sigma}=\frac{1}{2}(B_{\mu\rho,\sigma}+B_{\rho\sigma,\mu}+B_{\sigma\mu,\rho}$).}
are equal to zero \cite{Gat}. In the case that M is a Lie group G,
using non-coordinate basis, we have

\begin{equation}\label{11}
G_{\mu\nu}=L_{\mu}\hspace{0cm}^{A}L_{\nu}\hspace{0cm}^{B}G_{AB}=R_{\mu}\hspace{0cm}^{A}R_{\nu}\hspace{0cm}^{B}G_{AB},
\end{equation}
\vspace{-1mm}
\begin{equation}\label{13}
H_{\mu\nu\lambda}=\frac{1}{2}L_{\mu}\hspace{0cm}^{A}L_{\nu}\hspace{0cm}^{B}L_{\lambda}\hspace{0cm}^{C}H_{ABC}=\frac{1}{2}R_{\mu}\hspace{0cm}^{A}R_{\nu}
\hspace{0cm}^{B}R_{\lambda}\hspace{0cm}^{C}H_{ABC},
\end{equation}
\vspace{-1mm}
\begin{equation}\label{14}
J^{\mu}_{-\nu}=L^{\mu}\hspace{0cm}_{A}J^{A}\hspace{0cm}_{B}L_{\nu}\hspace{0cm}^{B},\hspace{5mm}J^{\mu}_{+\nu}=R^{\mu}\hspace{0cm}_{A}J^{A}\hspace{0cm}_{B}
R_{\nu}\hspace{0cm}^{B},
\end{equation}
\vspace{-.5mm} where $G_{AB}$ is symmetric ad-invariant
non-degenerate bilinear form (ad-invariant metric) and $H_{ABC}$
is antisymmetric tensor on Lie algebra $\bf g$; furthermore,
$L_{\mu}\hspace{0cm}^{A}(R_{\mu}\hspace{0cm}^{A})$ and
$L^{\mu}\hspace{0cm}_{A}(R^{\mu}\hspace{0cm}_{A})$ are left
(right) invariant veilbien \footnote{ Note that the indices
$A , B , ...$ show the Lie algebra indices and Greek indices $\mu
, \nu , ...$ show the Lie group manifold indices.} and their inverses on the Lie group $G$,
respectively, and $J^{A}\hspace{0cm}_{B}$ is an endomorphism of the Lie algebra ${\bf g}$; {\small{$J:~${\bf g}$
\longrightarrow${\bf g}$ $}}. Then, the conditions of the N=(2,2)
supersymmetric sigma model can be rewritten in the following
algebraic form \cite{RS}

\begin{equation}\label{15}
J^{2}=-I,\hspace{1mm}
\end{equation}
\vspace{-1mm}
\begin{equation}\label{16}
{\chi}_{A}+J^{t}\hspace{1mm}{\chi}_{A}\hspace{1mm}J^{t}+J^{B}\hspace{0cm}_{A}\hspace{1mm}{\chi}_{B}\hspace{1mm}J^{t}-J^{B}\hspace{0cm}_{A}\hspace{1mm}J^{t}
\hspace{1mm}{\chi}_{B}=0,
\end{equation}
\vspace{-1mm}
\begin{equation}\label{17}
J^{t}\hspace{1mm}G\hspace{1mm}J=G,
\end{equation}
\vspace{-1mm}
\begin{equation}\label{18}
H_{A}= J^{t} (H_{B} {J^{B}}\hspace{0cm}_{A}) + J^{t}H_{A}J+(H_{B}
{J^{B}}\hspace{0cm}_{A}) J,
\end{equation}
\vspace{-1mm}
\begin{equation}\label{19}
J^{t}(H_{A}+\chi_{A}G) =(J^{t}(H_{A}+\chi_{A}G))^{t},
\end{equation}
where  ${(\chi_A)_B}^C=-{f^C\hspace{0cm}_{AB}}$ is the adjoint representation
such that ${f^C\hspace{0cm}_{AB}}$ is the structure constant of the Lie
algebra $\bf{g}$ and we have $(H_A)_{BC}=H_{ABC}$,
furthermore, the ad-invariance of the metric implies that
\begin{equation}
\label{20} (\chi_A  G)^{t}=-\chi_A  G.
\end{equation}
These relations show that N=(2,2) supersymmetric sigma models on
the Lie groups from the geometric  point of view correspond
to the bi-Hermitian structures on the Lie groups \cite{Gat} or
equivalently the algebraic bi-Hermitian structures $(J,G, H)$ on
the Lie algebras \cite{RS}. For N=(2,2) supersymmetric WZW models on
the Lie group G, we have $H_{ABC}= f_{ABC}$. In this case,
relations (\ref{16}) and (\ref{17}) show that we have the Lie
bialgebra structures on $\bf g$ \cite{Par}\cite{Getzler}\cite{Lin}; and relation
(\ref{18}) reduces to (\ref{16}), and (\ref{19}) is automatically
satisfied, i.e., Lie bialgebra structure is a special case of
algebraic bi-Hermitian structure $(J,G,H)$ with $H_{ABC}=f_{ABC}$
\cite{RS}.

In \cite{ER}, we have considered the general cases such that the
perturbed N=(2,2) supersymmetric sigma model on a Lie group
preserve the N=(2,2) supersymmetry. Here, we assume that the action
(\ref{1}) (as sigma models on Lie groups) has the N=(2,2)
supersymmetry, and as a special example is perturbed with the
following term
\begin{equation}\label{22}
S^{'}=\int d^{2}\sigma d^{2}\theta D_{+}\Phi^{\mu}D_{-}\Phi^{\nu}
G_{\mu \lambda}(\Phi) {J^\lambda}_{\nu}(\Phi),
\end{equation}
where the tensor $J^{\lambda}\hspace{0cm}_{\nu}(\Phi)$
is the complex structure that appeared in the second supersymmetry transformations (\ref{3}) of the action (\ref{1}) . Now
together with (\ref{1}) we have
\begin{equation}\label{23}
S''=\int d^{2}\sigma d^{2}\theta
D_{+}\Phi^{\mu}D_{-}\Phi^{\nu}(G''_{\mu\nu}(\Phi)+B''_{\mu\nu}(\Phi)),
\end{equation}
such that
\begin{equation}\label{24}
 G''_{\mu\nu}=G_{\mu\nu}+\frac{1}{2}(G_{\mu\lambda}J^{\lambda}\hspace{0cm}_{\nu}+G_{\nu\lambda}J^{\lambda}\hspace{0cm}_{\mu}),
\end{equation}
\begin{equation}\label{25}
B''_{\mu\nu}=B_{\mu\nu}+\frac{1}{2}(G_{\mu\lambda}J^{\lambda}\hspace{0cm}_{\nu}-G_{\nu\lambda}J^{\lambda}\hspace{0cm}_{\mu}),
\end{equation}
with the inverse metric

\begin{equation}\label{26}
G''^{\mu\nu}=aG^{\mu\nu}+b(G^{\mu\lambda}J^{\nu}\hspace{0cm}_{\lambda}+G^{\nu\lambda}J^{\mu}\hspace{0cm}_{\lambda}).
\end{equation}
Now, using the condition $G'' G''^{-1}=1$ we have
\begin{equation}\label{r1}
a \delta^{\mu}\hspace{0cm}_{\nu} + (\frac{a}{2}+b) (J+A)^{\mu}\hspace{0cm}_{\nu}+\frac{b}{2}[(J+A)(J+A)]^{\mu}\hspace{0cm}_{\nu}=\delta^{\mu}\hspace{0cm}_{\nu}
\end{equation}
where $A=G^{-1}J^{t}G$. Therefore, we have different cases as follows:

\begin{equation}\label{r2}
case 1)~~~~~~~~~~~~~~~~J+A=0~~or~~G^{-1}J^{t}G=-J,
\end{equation}
which is satisfying for $a=1$ and all values of $b$,
\begin{equation}\label{r3}
case 2)~~~~~~~~~~~~~~J+A=I~~or~~G^{-1}J^{t}G=I-J,
\end{equation}
which is satisfying for $a+b=\frac{2}{3}$,
\begin{equation}\label{r4}
case 3)~~~~~~~~J+A=-I~~or~~G^{-1}J^{t}G=-(I+J),
\end{equation}
which is satisfying for $a-b=2$,
\begin{equation}\label{r5}
case 4)~~~~~~~~~(J+A)^{2}=I~~or~~(J+G^{-1}J^{t}G)^{2}=I,
\end{equation}
which is satisfying for $a=\frac{4}{3}$ and $b=-\frac{2}{3}$,
\begin{equation}\label{r6}
case 5)~~~~~~(J+A)^{2}=-I~~or~~(J+G^{-1}J^{t}G)^{2}=-I,
\end{equation}
which is satisfying for $a=\frac{4}{5}$ and $b=-\frac{2}{5}$. In the above cases, $I$ is identity matrix. Here, we choose the case$1)$ where $a=1$ and when
$J^{2}=-1$ then the Hermitian condition (\ref{17}) is satisfying such that the coefficient of the $b$  term in (\ref{26}) is automatically zero. Namely, $G''=G$ such that $S''$ and $S$ are different in $B$-field terms; i.e., $S$ is perturbed with complex structure (\ref{22}). For this selection, we have the {\bf{g)}} case (final case) in \cite{ER} where the algebraic bi-Hermitian structure $(J,G,H)$ is perturbed with $(0,0,H')$. Now, for having $S^{''}$ as a N=(2,2) supersymmetric sigma model, we must have the following relation (from relation (34) of \cite{ER})

\begin{equation}\label{27}
 H'_{A}J=(H'_{A}J)^{t},
\end{equation}
such that
\begin{equation}\label{28}
 H'_{ABC}=\frac{1}{2}({J^{D}}_{A}f_{DCB}+{J^{D}}_{B}f_{DAC}-{J^{D}}_{C}f_{DAB}).
\end{equation}
Now by substitution (\ref{28}) in (\ref{27}) and using
(\ref{15})-(\ref{17}) and (\ref{20}) we obtain

\begin{equation}\label{29}
\forall X,Y \in {\bf{g}} \hspace{10mm} J[X,Y]=[X,JY],
\end{equation}
where this relation with condition $J^2=-1$  shows that $J$ is an
{\it invariant complex structure} \cite{JO}. In this way, the N=(2,2)
supersymmetry is preserved in the action (\ref{23}) when the tensor
$J^{\lambda}\hspace{0cm}_{\nu}$ in (\ref{22}) is an invariant
complex structure \cite{JO}. Note that in \cite{JO}, it is proved
that Lie algebras with invariant complex structure compatible
with ad-invariant metric are the
Abelian Lie algebras. Therefore, only the Abelian Lie algebras admit this deformation which preserve the N=(2,2) supersymmetry.
\section{\bf Perturbed WZW model with complex structure as an integrable model}

In this section, we show that the perturbed bosonic WZW model with Hermitian complex structure (not necessarily invariant) is an integrable model. We know that the WZW model based on Lie group $G$  takes the
following standard form \cite{E} \vspace{-2mm}
$$
S_{WZW}(g) \; = \;  \frac{k}{4\pi} \int_{\Sigma} d\xi^{+} \wedge
d\xi^{-}\; < g^{-1} \partial_{+}g , \; g^{-1}
\partial_{-}g  >~~~~~~~~~~~~~
$$
\begin{equation}\label{30}
\;\;\; +\; \frac{k}{24\pi} \int_{B} < g^{-1}dg\; {,} \;[g^{-1}dg
\;{,}\;g^{-1}dg]
 >,\
\end{equation}
where the worldsheet $\Sigma$ is boundary
of 3-dimensional manifold $B$ $(\partial B = \Sigma)$,
and $g^{-1} d g $ with $g \in G$ is the left
invariant one-form on Lie group $G$ so that
\begin{equation}\label{31}
g^{-1} d g = (g^{-1} \partial_{\mu} g)^A X_{A} \partial_{\alpha}
x^{\mu} d\xi^{\alpha} = L_{\mu}\hspace{0cm}^{A} X_{A}
\partial_{\alpha} x^{\mu} d\xi^{\alpha},
\end{equation}
with $\{X_{A}\}$ as basis for the Lie algebra $ \bf g$ of the Lie
group G and $x^{\mu}$ are Lie group parameters. The WZW action (\ref{30}) then can be rewritten in the
following form
\begin{equation}\label{32}
S_{WZW}(g) =  \frac{k}{4\pi} \int_{\Sigma} d^2\xi\;
L_{\mu}\hspace{0cm}^{A}\;{G}_{AB}\;
L_{\nu}\hspace{0cm}^{B}\partial_{+} x^{\mu}\partial_{-}
x^{\nu}+\frac{k}{24\pi} \int_{B} d^3 \xi \epsilon^{
\alpha\beta\gamma} L_{\mu}\hspace{0cm}^{A} \;{G}_{AD}\;
L_{\nu}\hspace{0cm}^{B} {{f}^{D}}_{BC}\;
L_{\lambda}\hspace{0cm}^{C} \partial_{\alpha} x^{\mu}
\partial_{\beta} x^{\nu} \partial_{\gamma} x^{\lambda},
\end{equation}
where ${G}_{AB}=\;<X_A,X_B>$ is non-degenerate ad-invariant metric
on $\bf g$. Now, we assume that the action (\ref{32}) (as sigma
model on Lie group) is perturbed with the following term
\begin{equation}\label{34}
S^{'}= k^{'}\hspace{.5mm}\int\hspace{1mm} d\xi^{+} d\xi^{-}
(g^{-1}
\partial_{+}g)^{A} \hspace{2mm} G_{A D} {J^{D}}_{B} \hspace{2mm}(g^{-1}
\partial_{-}g)^{B},
\end{equation}
such that $J^{D}\hspace{0cm}_{B}$ is an endomorphism of ${\bf g}$,
i.e., $J: ${\bf g}$ \rightarrow ${\bf g}$ $. Now, using the
vielbein formalism \eqref{11} and \eqref{14} we have
\begin{equation}\label{38}
S^{'}= k^{'}\hspace{.5mm} \int\hspace{.5mm}d\xi^{+} d\xi^{-}
\hspace{.5mm} L_{\mu}\hspace{0cm}^{A}\hspace{1.5mm}
G_{AD}J^{D}\hspace{0cm}_{B}\hspace{1.5mm}
L_{\nu}\hspace{0cm}^{B}\hspace{.5mm}{\partial}_{+}\hspace{0cm}x^{\mu}{\partial}_{-}x^{\nu}=
k^{'} \int\hspace{.5mm}d\xi^{+}
d\xi^{-}\hspace{.5mm}G_{\mu\lambda}J^{\lambda}\hspace{0cm}_{\nu}\hspace{.5mm}{\partial}_{+}\hspace{0cm}x^{\mu}{\partial}_{-}x^{\nu}.
\end{equation}
Then the general action of the WZW model where perturbed with the term
(\ref{34}) can be rewritten as follows

\begin{equation}\label{39}
S^{''} \equiv\ S+S^{'}=\int\hspace{.5mm}d\xi^{+} d\xi^{-}
\hspace{.5mm} L_{\mu}\hspace{0cm}^{A}(G_{AB}+B_{AB}+
k^{'}\hspace{.5mm}G_{AD}J^{D}\hspace{0cm}_{B})L_{\nu}\hspace{0cm}^{B}\hspace{.5mm}{\partial}_{+}\hspace{0cm}x^{\mu}{\partial}_{-}x^{\nu},
\end{equation}
such that $S^{''}$ as sigma model have the following invertible metric and anti-symmetric tensor (by choosing $k=4\pi$)\\
$$G^{''}_{\mu \nu}=L_{\mu}\hspace{0cm}^{A} G_{AB} L_{\nu}\hspace{0cm}^{B}\hspace{.5mm}+ \frac {k^{'}} {2}(L_{\mu}\hspace{0cm}^{A}G_{AD}J^{D}\hspace{0cm}_{B}L_{\nu}\hspace{0cm}^{B}+L_{\nu}\hspace{0cm}^{A}G_{AD}J^{D}\hspace{0cm}_{B}L_{\mu}\hspace{0cm}^{B}),$$
\begin{equation}\label{40}
~~B^{''}_{\mu \nu}=L_{\mu}\hspace{0cm}^{A} B_{AB}
L_{\nu}\hspace{0cm}^{B}\hspace{.5mm}+ \frac {k^{'}}
{2}(L_{\mu}\hspace{0cm}^{A}G_{AD}J^{D}\hspace{0cm}_{B}L_{\nu}\hspace{0cm}^{B}-L_{\nu}\hspace{0cm}^{A}G_{AD}J^{D}\hspace{0cm}_{B}L_{\mu}\hspace{0cm}^{B}),
\end{equation}
with the inverse metric
\begin{equation}\label{41}
{G^{''}}^{\mu \nu}=a\hspace{.5mm}L^{\mu}\hspace{0cm}_{A} G^{AB}
L^{\nu}\hspace{0cm}_{B}\hspace{.5mm}+ b\hspace{.5mm}
(L^{\mu}\hspace{0cm}_{A}G^{AD}J^{B}\hspace{0cm}_{D}L^{\nu}\hspace{0cm}_{B}+L^{\nu}\hspace{0cm}_{A}G^{AD}J^{B}\hspace{0cm}_{D}L^{\mu}\hspace{0cm}_{B}).
\end{equation}
Similar to the previous section using $G^{''} {G^{''}}^{-1}=1$ and
Hermitian condition $J^{t} G^{''} J= G^{''}$ we will arrive at
$a=1$ such that the coefficient of the $b$ term must be zero in
${G^{''}}^{\mu \nu}$. Now, we will prove that the model (\ref{39}) is an
integrable model. For this purpose, we use the formalism presented
in \cite{MO}. In this direction, we should calculate
$H^{\lambda}\hspace{0cm}_{\mu \nu}$,
$\Gamma^{\lambda}\hspace{0cm}_{\mu \nu}$ and
$\Omega^{\lambda}\hspace{0cm}_{\mu \nu}$; so after some
calculation we have
\begin{equation}\label{45}
\Gamma^{\lambda}\hspace{0cm}_{\mu \nu}=\frac{1}{2}({-f^{A}}_{B C}
+2
(\partial_{\rho}L_{\sigma}\hspace{0cm}^{A})L^{\sigma}\hspace{0cm}_{B}L^{\rho}\hspace{0cm}_{C})L^{\lambda}
\hspace{0cm}_{A}L_{\mu}\hspace{0cm}^{B}L_{\nu}\hspace{0cm}^{C},~~~~~~~~~~~~~~~~~~~~~~~~~~~~
\end{equation}
\begin{equation}\label{46}
~~~~{H^{''}}^{\lambda}\hspace{0cm}_{\mu\nu}=
\frac{1}{2}(H^{A}\hspace{0cm}_{BC}+k^{'}\hspace{.5mm}
J^{D}\hspace{0cm}_{B}\hspace{1mm}f^{A}\hspace{0cm}_{C D}-
k^{'}\hspace{.5mm}J^{A}\hspace{0cm}_{D}\hspace{1mm}f^{D}\hspace{0cm}_{C
B}-k^{'}\hspace{.5mm}
J^{D}\hspace{0cm}_{C}\hspace{1mm}f^{A}\hspace{0cm}_{B
D})L^{\lambda}\hspace{0cm}_{A}L_{\mu}\hspace{0cm}^{B}L_{\nu}\hspace{0cm}^{C},
\end{equation}
$$
\Omega^{\lambda}\hspace{0cm}_{\mu\nu}=\Gamma^{\lambda}\hspace{0cm}_{\mu \nu}-{H^{''}}^{\lambda}\hspace{0cm}_{\mu\nu}
=\frac{1}{2}(-H^{A}\hspace{0cm}_{BC}-f^{A}\hspace{0cm}_{BC}-k^{'}\hspace{.5mm}
J^{D}\hspace{0cm}_{B}\hspace{1mm}f^{A}\hspace{0cm}_{C D}+
k^{'}\hspace{.5mm}J^{A}\hspace{0cm}_{D}\hspace{1mm}f^{D}\hspace{0cm}_{C
B}~~~~~~~~~~~
$$
\begin{equation}\label{47}
~~~~~~~~~~~~~~~~+
k^{'}\hspace{.5mm}J^{D}\hspace{0cm}_{C}\hspace{1mm}f^{A}\hspace{0cm}_{B
D}+2
(\partial_{\rho}L_{\sigma}\hspace{0cm}^{A})L^{\sigma}\hspace{0cm}_{B}L^{\rho}\hspace{0cm}_{C})L^{\lambda}\hspace{0cm}_{A}L_{\mu}\hspace{0cm}^{B}L_{\nu}
\hspace{0cm}^{C},
\end{equation}
where in this calculation we use
\begin{equation}\label{49}
H_{\mu\nu\lambda}=\frac{1}{2}(\partial_{\lambda}B_{\mu\nu}+\partial_{\nu}B_{\lambda\mu}+\partial_{\mu}B_{\nu\lambda})=\frac{1}{2}
L_{\mu}\hspace{0cm}^{A} L_{\nu}\hspace{0cm}^{B}
L_{\lambda}\hspace{0cm}^{C} H_{A B C},
\end{equation}
such that for the WZW model we have
\begin{equation}\label{48}
H_{A B C}=f_{A B C}.
\end{equation}
In this way, we must obtain a Lax pair as follows\cite{MO}
$$
[\partial_{+}+\alpha_{\mu}(x)\partial_{+}\hspace{0cm}x^{\mu}]\psi=0,
$$
\begin{equation}\label{50}
[{\partial}_{-}+\beta_{\nu}(x)\partial_{-}x^{\nu}]\psi=0,
\end{equation}
where, consistency condition (the zero curvature condition) of this linear system yields the
equation of motion of the system (\ref{39}), such that $\psi$ is arbitrary function of the fields $x^{\mu}$; and the matrices $\alpha_{\mu}(x)$ and
$\beta_{\mu}(x)$ satisfy  the following relations \cite{MO}
\begin{equation}\label{51}
\beta_{\mu}-\alpha_{\mu}=\mu_{\mu},
\end{equation}
\begin{equation}\label{52}
 \partial\hspace{0cm}_{\mu}\beta_{\nu}-\partial\hspace{0cm}_{\nu}\alpha_{\mu}+[\alpha_{\mu},\beta_{\nu}]=\Omega^{\lambda}\hspace{0cm}_{\mu\nu}
 \mu_{\lambda},
\end{equation}
where equation (\ref{52}) can then be rewritten as \cite{MO}
\begin{equation}\label{53}
 F_{\mu\nu}=-(\nabla_{\mu}\mu_{\nu}-\Omega^{\lambda}\hspace{0cm}_{\mu\nu}\mu_{\lambda}),
\end{equation}
so that the field strength $F_{\mu\nu}$ and covariant derivative
are written as follows
\begin{equation}\label{55}
 F_{\mu\nu}=\partial_{\mu}\alpha_{\nu}-\partial_{\nu}\alpha_{\mu}+[\alpha_{\mu},\alpha_{\nu}],\hspace{1cm}
 \nabla_{\mu}X=\partial_{\mu}X+[\alpha_{\mu},X].
\end{equation}
Now for our model (\ref{39}) we choose
\begin{equation}\label{56}
\alpha_{\mu}=c~C^{A}\hspace{0cm}_{B}L_{\mu}\hspace{0cm}^{B}X_{A},\hspace{1cm}\mu_{\mu}=d~D^{A}\hspace{0cm}_{B}L_{\mu}\hspace{0cm}^{B}X_{A},
\end{equation}
where c, d are constant and $C^{A}\hspace{0cm}_{B}$, $D^{A}\hspace{0cm}_{B}$ are constant matrices. By assuming that $G_{AB}$ and
$J^{A}\hspace{0cm}_{B}$ are independent of the coordinate of the
Lie group G; after some calculation we see that for satisfying
relation (\ref{53}) we must have the following relation
\begin{equation}\label{57}
C^{A}\hspace{0cm}_{B}=J^{A}\hspace{0cm}_{B}+\delta^{A}\hspace{0cm}_{B},\hspace{5mm}D^{A}\hspace{0cm}_{B}=J^{A}\hspace{0cm}_{B}+\delta^{A}\hspace{0cm}_{B},
\end{equation}
with
\begin{equation}\label{58}
c=\frac{k^{'}\hspace{.5mm}}{{k^{'}}\hspace{.5mm}+2},~~~~~d=\frac{2}{{k^{'}}\hspace{.5mm}+2}\hspace{1mm},
\end{equation}
such that $J^{A}\hspace{0cm}_{B}$ must satisfy the
(\ref{15})-(\ref{17}), i.e., it must be an {\it algebraic Hermitian complex
structure}. In this way, we show that the WZW model (\ref{32})
which is perturbed with (\ref{34}), is integrable when $J$ is a
Hermitian complex structure (not necessarily invariant) on the Lie algebra $\bf{g}$. Furthermore, the
equations of motion for this integrable model can be rewritten as
the following Lax pair
$$
[\partial_{+}+\frac{k^{'}\hspace{.5mm}}{{k^{'}}\hspace{.5mm}+2}\;(J^{A}\hspace{0cm}_{B}+\delta^{A}\hspace{0cm}_{B})L_{\mu}\hspace{0cm}^{B}
X_{A}\partial_{+}x^{\mu}]\psi=0,
$$
\begin{equation}\label{50}
[{\partial}_{-}+\;(J^{A}\hspace{0cm}_{B}+\delta^{A}\hspace{0cm}_{B})L_{\nu}\hspace{0cm}^{B}X_{A}\partial_{-}x^{\nu}]\psi=0.
\end{equation}
But the presence of a spectral parameter in the Lax pair is of crucial importance in extracting conserved quantities \cite{F},\cite{B}\footnote{In the Lax formulation of classical integrability of a two-dimensional field theory (in general) \cite{F} and a two-dimensional sigma model (as a special case) \cite{Ph}, the equations of motion for the system can be written as a consistency condition (the zero curvature condition) in the following linear system
\begin{equation}\label{f1}
[\partial_{0}+{\cal A}_{0}({\lambda})]\psi=0,
\end{equation}
\begin{equation}\label{f2}
[\partial_{1}+{\cal A}_{1}({\lambda})]\psi=0,
\end{equation}
where $\partial_{0}=\frac{\partial}{\partial \tau}$ , $\partial_{1}=\frac{\partial}{\partial \sigma}$ and ${\cal A}_{0},{\cal A}_{1}$ are matrices which depend on the fields $x^{\mu}$ and the arbitrary spectral parameter $\lambda$, such that the zero curvature condition is as follows
\begin{equation}\label{f3}
\{\partial_{0}{\cal A}_{1}-\partial_{1}{\cal A}_{0}+[{\cal A}_{0},{\cal A}_{1}]\}\psi=0.
\end{equation}
For such systems, the monodromy matrix has the following form
\begin{equation}\label{f4}
T(\lambda,\tau)=P \exp(-\int_{0}^{2\pi} {\cal A}_{1}(\lambda, \sigma, \tau) d\sigma),
\end{equation}
where $P$ represents the path-ordered exponential. Since, traces of powers of the monodromy matrix are independent of time (for the proof of this statement, the assumption of the periodicity condition ${\cal A}_{0}(\lambda,0,\tau)={\cal A}_{0}(\lambda,2\pi,\tau)$ is necessary \cite{B}), the conserved quantities can be obtained from the following relation \cite{B}
\begin{equation}\label{f5}
H^{n}(\lambda)=Tr[T^{n}(\lambda,\tau)].
\end{equation}
These quantities are independent of time $\tau$ and in involution with respect to the Poisson brackets: $\{H^{n}(\lambda_{1}),H^{m}(\lambda_{2})\}=0$ (one can find their proofs in \cite{F},\cite{B}). In this paper, we use the light cone coordinates $\xi^{\pm}=\frac{\tau\pm\sigma}{2}$ instead of $(\tau,\sigma)$.}. For this purpose, we will prove that $k'$ parameter in (\ref{39}) plays the role of the spectral parameter. In order to show this, we must obtain a Lax pair as follows \cite{MO}
$$
[\partial_{+}+y\hat{\alpha}_{\mu}(x)\partial_{+}\hspace{0cm}x^{\mu}]\psi=0,
$$
\begin{equation}\label{r7}
[{\partial}_{-}+z(y)\hat{\beta}_{\nu}(x)\partial_{-}x^{\nu}]\psi=0,
\end{equation}
where $y$ is a multiplicative spectral parameter and $z(y)$ is a function of $y$. The matrices $\hat{\alpha}_{\mu}$, $\hat{\beta}_{\nu}$ are independent of $y$.
The equivalence of the consistency condition of the linear system and the equations of motion of the nonlinear sigma model results as following equations
\begin{equation}\label{r8}
z\hat{\beta}_{\mu}-y\hat{\alpha}_{\mu}=\hat{\mu}_{\mu},
\end{equation}
\begin{equation}\label{r9}
 z\partial\hspace{0cm}_{\mu}\hat{\beta}_{\nu}-y\partial\hspace{0cm}_{\nu}\hat{\alpha}_{\mu}+yz[\hat{\alpha}_{\mu},\hat{\beta}_{\nu}]=\Omega^{\lambda}
 \hspace{0cm}_{\mu\nu}
 \hat{\mu}_{\lambda},
\end{equation}
where equation (\ref{r9}) in the terms of matrices $\hat{\alpha}_{\mu},\hat{\beta}_{\nu}$ is rewritten as follows
\begin{equation}\label{r10}
z(\partial\hspace{0cm}_{\mu}\hat{\beta}_{\nu}-\Omega^{\lambda}\hspace{0cm}_{\mu\nu}
 \hat{\beta}_{\lambda})-y(\partial\hspace{0cm}_{\nu}\hat{\alpha}_{\mu}-\Omega^{\lambda}\hspace{0cm}_{\mu\nu}
 \hat{\alpha}_{\lambda})+yz[\hat{\alpha}_{\mu},\hat{\beta}_{\nu}]=0.
\end{equation}
Now, for our model (\ref{39}) we choose
\begin{equation}\label{r11}
\hat{\alpha}_{\mu}=(J^{A}\hspace{0cm}_{B}+\delta^{A}\hspace{0cm}_{B})L_{\mu}\hspace{0cm}^{B}X_{A},\hspace{1cm}
\hat{\beta}_{\nu}=(J^{A}\hspace{0cm}_{B}+\delta^{A}\hspace{0cm}_{B})L_{\nu}\hspace{0cm}^{B}X_{A},
\end{equation}
Using (\ref{47}), by assuming that $G_{AB}$ and
$J^{A}\hspace{0cm}_{B}$ are independent of the coordinate of the
Lie group G, after some calculation we see that for satisfying
relation (\ref{r10}) we must have the following relations
\begin{equation}\label{r12}
y=\frac{k^{'}\hspace{.5mm}}{{k^{'}}\hspace{.5mm}+2},~~~~~z=1.
\end{equation}
By substituting the above results in the equations of (\ref{r7}), we will obtain the previous Lax pair (\ref{50}).
Therefore, $y=\frac{k^{'}\hspace{.5mm}}{{k^{'}}\hspace{.5mm}+2}$ is a spectral parameter where by this parameter the model (\ref{39}) is an integrable sigma model. In order to clarify this integrability, we will read the general form of conserved quantities for this model as follows. First, we express the equation of motion (\ref{50}) in the terms of $(\tau,\sigma)$ coordinates, then we will have Lax pair $({\cal A}_{0},{\cal A}_{1})$ in the following forms
\begin{equation}\label{r13}
{\cal A}_{0}(y)=(J^{A}\hspace{0cm}_{B}+\delta^{A}\hspace{0cm}_{B})L_{\mu}\hspace{0cm}^{B}X_{A}(\frac{y+1}{2}\partial_{0}x^{\mu}+\frac{y-1}{2}\partial_{1}x^{\mu}),
\end{equation}
\begin{equation}\label{r14}
{\cal A}_{1}(y)=(J^{A}\hspace{0cm}_{B}+\delta^{A}\hspace{0cm}_{B})L_{\mu}\hspace{0cm}^{B}X_{A}(\frac{y-1}{2}\partial_{0}x^{\mu}+\frac{y+1}{2}\partial_{1}x^{\mu}).
\end{equation}
Using (\ref{f4}) and (\ref{r14}), the monodromy matrix for model (\ref{39}) (by assuming that $J^{A}\hspace{0cm}_{B}$ is independent of the coordinates $(\tau,\sigma)$) has the following form
\begin{equation}\label{r15}
T(y,\tau)=P \exp{\big\{-}(J^{A}\hspace{0cm}_{B}+\delta^{A}\hspace{0cm}_{B})X_{A}(\frac{y-1}{2}\int_{0}^{2\pi}L_{\mu}\hspace{0cm}^{B}\partial_{0}x^{\mu}d\sigma
+\frac{y+1}{2}\int_{0}^{2\pi}L_{\mu}\hspace{0cm}^{B}\partial_{1}x^{\mu}d\sigma){\big\}}.
\end{equation}
According to the above relation for monodromy matrix, conserved quantities mentioned in (\ref{f5}) are dependent on spectral parameter $y$. Indeed, the monodromy matrix (\ref{r15}) of our model (\ref{39}) completely coincides with that of WZW model \cite{MO}\cite{ABD1}\cite{ABD2} with this difference that in our model $\delta^{A}\hspace{0cm}_{B}$ is replaced with $J^{A}\hspace{0cm}_{B}+\delta^{A}\hspace{0cm}_{B}$\footnote{In the light cone coordinates $\xi^{\pm}=\frac{\tau\pm\sigma}{2}$ the equations of motion for WZW model are written in the following Lax form \cite{MO}
$$
[\partial_{+}+x\;L_{\mu}\hspace{0cm}^{A} X_{A}\partial_{+}x^{\mu}]\psi=0,
$$
\begin{equation}\label{r16}
[{\partial}_{-}+\;L_{\nu}\hspace{0cm}^{A} X_{A}\partial_{-}x^{\nu}]\psi=0.
\end{equation}
and Lax pair $({\cal A}_{0},{\cal A}_{1})$ are in the following forms
\begin{equation}\label{r17}
{\cal A}_{0}(x)=L_{\mu}\hspace{0cm}^{A}X_{A}(\frac{x+1}{2}\partial_{0}x^{\mu}+\frac{x-1}{2}\partial_{1}x^{\mu}),
\end{equation}
\begin{equation}\label{r18}
{\cal A}_{1}(x)=L_{\mu}\hspace{0cm}^{A}X_{A}(\frac{x-1}{2}\partial_{0}x^{\mu}+\frac{x+1}{2}\partial_{1}x^{\mu}).
\end{equation}
Therefore, we have the monodromy matrix for WZW model as follows
\begin{equation}\label{r19}
T(x,\tau)=P \exp{\big\{-}X_{A}(\frac{x-1}{2}\int_{0}^{2\pi}L_{\mu}\hspace{0cm}^{A}\partial_{0}x^{\mu}d\sigma
+\frac{x+1}{2}\int_{0}^{2\pi}L_{\mu}\hspace{0cm}^{A}\partial_{1}x^{\mu}d\sigma){\big\}},
\end{equation}
where $x$ is a spectral parameter.}}. From expanding $H^{n}(y)$ in $y$ we obtain an infinite set of conserved quantities. This suggests that model (\ref{39}) is an integrable field theory\cite{B}.

\subsection{\bf An example}

In the following, we will give example for WZW model perturbed with
complex structure on Heisenberg Lie group ${\bf{A_{4,8}}}$\footnote{Note that in real domain this Lie algebra is different from the Nappi-Witten Lie algebra \cite{NW} which is isomorphic to ${\bf{A_{4,10}}}$ \cite{P}; however, in complex domain they are isomorphic.}.
Heisenberg Lie algebra with basis $\{X_{1},..,X_{4}\}$ has the
following set of non-trivial commutation relations \cite{P}
\begin{equation}\label{59}
[X_{2},X_{4}]=X_{2} ,\hspace{1cm}[X_{3},X_{4}]=-X_{3}
,\hspace{1cm}[X_{2},X_{3}]=X_{1}.
\end{equation}
Here, we obtain a non-degenerate ad-invariant metric by using the
general solution of (\ref{20}) as follows
\begin{eqnarray}\label{60}
G_{AB}=\left(
              \begin{array}{cccc}
                0 & 0 & 0 & -a \\
                0 & 0 & a & 0 \\
                0 & a & 0 & 0 \\
                -a & 0 & 0 & b \\
              \end{array}
              \right),\qquad   a\in \Re-\{0\},\;\;\; b\in \Re.
\end{eqnarray} In order to write (\ref{30}) and (\ref{34})
explicitly, we need $g^{-1}\partial_{\alpha} g$. To this end, we
use the following parametrization of Lie group G
\begin{equation}\label{61}
g\;=\; e^{x^{1} X_{1}}  e^{x^{2} X_{2}}  e^{x^{3} X_{3}} e^{x^{4}
X_{4}},
\end{equation}
where ${X_{i}}$ and ${x^{i}}$ are generators and parameters of
Lie group $G$, respectively. Inserting our specific choice of the
parametrization (\ref{61}) then $g^{-1}\partial_{\alpha} g$ takes
the following form \cite{MR}
\begin{equation}\label{62}
g^{-1}\partial_{\alpha} g=(\partial_{\alpha}x^{1})
X_{1}+(\partial_{\alpha}x^{2})(x^{3}X_{1}+e^{x^{4}}X_{2})+(\partial_{\alpha}x^{3})(e^{-x^{4}}X_{3})+(\partial_{\alpha}x^{4})
X_{4},
\end{equation}
from which we can read the ${L_{\mu}}\hspace{0cm}^{A}$'s and
the terms that are being integrated over in (\ref{32}) as follows \footnote{Note that we choose
$\epsilon_{\alpha\beta}=\left(
              \begin{array}{cc}
                0 & -1 \\
                1 & 0 \\
              \end{array}\right)$ in light cone coordinate.}
\begin{equation}\label{63}
L_{+}\hspace{0cm}^{A}\;{G}_{AB}\; L_{-}\hspace{0cm}^{B} =
[\partial_{+}x^{1}\partial_{-}x^{4}+\partial_{+}x^{4}\partial_{-}x^{1}
-\partial_{+}x^{2}\partial_{-}x^{3}-\partial_{+}x^{3}\partial_{-}x^{2}+x^{3}\partial_{+}x^{2}\partial_{-}x^{4}+x^{3}\partial_{+}x^{4}\partial_{-}x^{2}],
\end{equation}
\begin{equation}\label{64}
\epsilon^{ \alpha\beta\gamma} L_{\alpha}\hspace{0cm}^{A}
\;{G}_{AD}\; L_{\beta}\hspace{0cm}^{B} ({\cal{Y}}^D)_{BC}\;
L_{\gamma}\hspace{0cm}^{C} = -2 \epsilon^{
\alpha\beta\gamma}\partial_{\gamma}
[x^{3}\partial_{\alpha}x^{4}\partial_{\beta}x^{2}-x^{4}\partial_{\alpha}x^{3}\partial_{\beta}x^{2}-x^{2}\partial_{\alpha}x^{4}\partial_{\beta}x^{3}],
\end{equation}
where in the above relations we choose $b=0$, $a=-1$ in
(\ref{60}), and use the adjoint representation $({\cal{Y}}^l)_{jk}
= -{{f}}^{l}\hspace{0cm}_{jk}$ and also use the following relation
\begin{equation}\label{65} L_{\alpha}\;\equiv\;g^{-1}
\partial_{\alpha}g\;=\;(g^{-1}\partial_{\alpha}g)^A X_{A}=L_{\mu}\hspace{0cm}^{A} X_{A}\partial_{\alpha}x^{\mu}.
\end{equation}
On the other hand, using integration by parts, the action
(\ref{32}) is reduced to
\begin{equation}\label{66}
S_{WZW}(g) =  \frac{k}{2\pi} \int_{\Sigma} d^2\xi\
(\partial_{+}x^{1}\partial_{-}x^{4}
-\partial_{+}x^{2}\partial_{-}x^{3}+x^{3}\partial_{+}x^{4}\partial_{-}x^{2}).
\end{equation}
Now, for calculation of the perturbed term (\ref{34}) we first
choose complex structure compatible with metric (\ref{60}) in which
$b=0$ , $a=-1$; from \cite{RS} we have
\begin{eqnarray}\label{67}
J=\left(
              \begin{array}{cccc}
                0 & -1 & 0 & 0 \\
                1 & 0 & 0 & 0 \\
                0 & 0 & 0 & -1 \\
                0 & 0 & 1 & 0 \\
              \end{array}
              \right)
,\end{eqnarray} by this selection, the action (\ref{34}) takes the
following form
\begin{equation}\label{68}
S^{'}= k^{'} \int\hspace{1mm} d^2\xi\
(e^{-x^{4}}{\partial_{+}x^{1}}{\partial_{-}x^{3}}-e^{-x^{4}}{\partial_{+}x^{3}}{\partial_{-}x^{1}}+x^{3}e^{-x^{4}}\partial_{+}x^{2}\partial_{-}x^{3}
-x^{3}e^{-x^{4}}\partial_{+}x^{3}\partial_{-}x^{2}).
\end{equation}
Finally, perturbed WZW model (\ref{39}) by choosing $k=4\pi$ takes
the following integrable model
$$ S^{''}= \int\hspace{1mm} d^2\xi\ [k^{'}e^{-x^{4}}{\partial_{+}x^{1}}{\partial_{-}x^{3}}+2\partial_{+}x^{1}\partial_{-}x^{4}
-(1-k^{'}x^{3}e^{-x^{4}})\partial_{+}x^{2}\partial_{-}x^{3}
$$
\begin{equation}\label{69}
~~~~~~~~~~~~~~~-k^{'}e^{-x^{4}}{\partial_{+}x^{3}}{\partial_{-}x^{1}}-(1+k^{'}x^{3}e^{-x^{4}})\partial_{+}x^{3}\partial_{-}x^{2}
+2x^{3}\partial_{+}x^{4}{\partial_{-}x^{2}}],
\end{equation}
such that the equations of motion can be rewritten in the
following Lax form
$$
[\partial_{+}+\frac{k^{'}\hspace{.5mm}}{{k^{'}}\hspace{.5mm}+2}\{\;(\partial_{+}x^{1}+(x^3-e^{x^4})\partial_{+}x^{2})X_1+(\partial_{+}x^{1}+(x^{3}+e^{x^4})
\partial_{+}x^{2})X_2$$
\begin{equation}\label{70}
~~~~~~~~~~~~~~~~~+(e^{-x^4}\partial_{+}x^{3}-\partial_{+}x^{4})X_3+(e^{-x^4}\partial_{+}x^{3}+\partial_{+}x^{4})X_4\}]\psi=0,
\end{equation}

$$[\partial_{-}+
\{\;(\partial_{-}x^{1}+(x^3-e^{x^4})\partial_{-}x^{2})X_1+(\partial_{-}x^{1}+(x^{3}+e^{x^4})\partial_{-}x^{2})X_2
$$
\begin{equation}\label{71}
~~~~~~~~+(e^{-x^4}\partial_{-}x^{3}-\partial_{-}x^{4})X_3+(e^{-x^4}\partial_{-}x^{3}+\partial_{-}x^{4})X_4\}]\psi=0.
\end{equation}
\section{\bf Conclusion}

We have proved that N=(2,2) supersymmetric sigma models on Lie groups when perturbed
with complex structure can preserve the N=(2,2) supersymmetry if their Lie algebras have invariant
complex structure compatible with ad-invariant metric, i.e., to be the Abelian Lie algebras. Also,
we have shown that the zero curvature representation or consistency of integrability condition for
bosonic WZW models perturbed with this term is equivalent to the vanishing of the Nijenhuis tensor
for the Hermitian complex structure (not necessarily invariant).\\

 {\it{Acknowledgment}}: We would like to thank F. Darabi and A. Eghbali for careful reading of the paper and their useful comments. This research was supported by a research fund No. 217/D/1639 from Azarbaijan Shahid Madani University.


\end{document}